\documentclass[12pt]{article}

\newif\ifpdf
\ifx\pdfoutput\undefined
\pdffalse 
\else
\pdfoutput=1 
\pdftrue
\fi
 
\ifpdf
\usepackage[pdftex]{graphicx}
\else
\usepackage{graphicx}
\fi

\def\bit{\begin{itemize}}                                                      
\def\eit{\end{itemize}}
 \newcommand{\note}[1]{}

\newcommand{\delhad}{\mbox{$\Delta \alpha_{\rm had}^{(5)}(M_Z)$}} 
\newcommand{\msb}{\mbox{$\overline{\rm{MS}}\ $}}                                
\newcommand{\mt}{\mbox{$m_t$}}                                                  
\newcommand{\mh}{\mbox{$M_H$}}

\newcommand{\als}{\mbox{$\alpha_s$}}

\newcommand{\skipblk}[1]{}                                                      
\def\bqa{\begin{eqnarray}}                                                      
\def\eqa{\end{eqnarray}}

\newcommand{\x}{\mbox{$\times$}}

\newcommand{\sinn}{\mbox{$\sin^2\theta_W\,$}}                                   
                                            
\newcommand{\snu}{\mbox{$\stackrel{(-)}{\nu}$}}                                 
\newcommand{\beq}{\begin{equation}}                                             
\newcommand{\eeq}{\end{equation}}

\newcommand{\RA}{\mbox{$\rightarrow$}}

\def\mxth{\mathsurround=0pt }
\def\xversim#1#2{\lower2.pt\vbox{\baselineskip0pt \lineskip-.5pt
  \ialign{$\mxth#1\hfil##\hfil$\crcr#2\crcr\sim\crcr}}}             
\def\simgr{\mathrel{\mathpalette\xversim >}}                                    
\def\simle{\mathrel{\mathpalette\xversim <}}

\begin{document}

 \ifpdf
\DeclareGraphicsExtensions{.jpg,.pdf,.mps,.png}
 \else
\DeclareGraphicsExtensions{.eps,.ps}
 \fi


\title{Recent Developments in  Precision Electroweak
Physics\footnote{This is an update on recent developments,
prepared for the publication of the Proceedings of
Alberto Sirlin Symposium, New York University, October 2000.}}
\author{Paul Langacker\\
Department of Physics and Astronomy \\
University of Pennsylvania \\
Philadelphia, PA 19104}
\maketitle
\abstract{Developments in precision electroweak physics in the
two years since the symposium are briefly summarized.}

\section{Introduction}
The precision electroweak program has been remarkably successful in
demonstrating: (a) that the standard electroweak model is correct to first approximation,
verifying the gauge principle, the $SU(2) \times U(1)$ group, and the representations;
(b) that the QED, electroweak, QCD, and mixed 
radiative corrections are correct, verifying
the validity of renormalizable gauge theories; (c) determining such standard model
parameters as the weak angle and predicting $m_t$, \als, and \mh; and (d)
severely constraining posssible new physics at the TeV scale to be of the decoupling
type, such as supersymmetry. All of these 
issues were thoroughly explored in the talks and
written versions of the Alberto-Fest, held at NYU in October 2000. It is the
purpose of this note to  mention a few of the major developments in the
subsequent two years. More details may be found in~\cite{pdg,lepewwg,snowmass}.

\section{New inputs and anomalies}
There have been a number of recent important results, some hinting at the
possibility of new physics.

\begin{itemize}
\item The direct determination of the $W$ mass from CDF, D\O, and UA2,
currently $80.454(59)$ GeV~\cite{lepewwg,je} has for some time been on the high side of
the standard model fit prediction of $80.391(18)$ GeV~\cite{je}. However, the recent
LEP2 determination of $80.450(39)$~\cite{lepewwg} is also slightly high, leading
to a combined value of $80.451(33)$ GeV. This is only $1.8\sigma$ above the best fit
prediction, but  contributes to the small predicted value for the Higgs mass.

\item Preliminary analysis of the LEP2 data at energies in excess of
206 GeV indicated evidence for the Higgs boson at around 115 GeV.
The significance of the signal has been considerably reduced in the
final analyses~\cite{lephiggs}, although there is still a hint, especially
in the ALEPH four-jet data~\cite{ALEPH}. The LEP Higgs working group
now quotes a preliminary combined lower limit $M_H > 114.4$ GeV at 95\% cl
on the standard model Higgs.

\item There is a new  estimate of \als \ from the $\tau$
lifetime~\cite{je,tauwidth}, which is quite precise though theory-error dominated,
yielding \als$(M_\tau) = 0.356^{+0.027}_{-0.021}$, corresponding
to \als$(M_Z) = 0.1221^{+0.0026}_{-0.0023}$.

\item $A_{FB}(b)$, the forward-backward asymmetry into $b$ quarks, has the
value $0.0994(17)$, which is 2.6$\sigma$ below the standard model global
fit value of 0.1038(8). On the other hand, the SLD value for the related quantity
$A_b= 0.922(20)$ (see the appendices in~\cite{pl} for the definitions)
is only  $0.6\sigma$ below the expected 0.9347(1), and
the hadronic branching fraction $R_b=0.2165(7)$, which at one time
appeared anomalous, is now only 1.1$\sigma$ above the expectation 0.2157(2). 
If not just a statistical fluctuation or systematic problem, $A_{FB}(b)$ could
be a hint of new physics.  However, any such effect should not contribute
too much to  $R_b$. The deviation is only around 5\%, but if the
new physics involved a 
 radiative correction to the coefficient $\kappa$ \
of \sinn, the change would have to be around
25\%. Hence, the new physics would most likely be at the
tree level, mainly increasing the magnitude of the right-handed coupling
to the $b$. This could be due to a heavy $Z'$ boson
with non-universal couplings to the third family~\cite{zpr,fcnc};
or to the mixing of the $b_R$ with exotic quarks~\cite{pdg,wagner},
such as with an $SU(2)$ doublet involving a heavy $B_R$ quark and a charge
$-4/3$ partner~\cite{wagner}.
There is a strong correlation between $A_{FB}(b)$ and
the predicted Higgs mass $M_H$ in the global fits. It
has been emphasized~\cite{chanowitz} that if one eliminated $A_{FB}(b)$ from the
fit (e.g., because it is affected by new physics) then the $M_H$
prediction would be lower, with the central value well below the
lower limit from the direct searches at LEP2.
One resolution, assuming $A_{FB}(b)$  is due to an experimental
problem or fluctuation, is to invoke a supersymmetric extension
of the standard model with light sneutrinos, sleptons, and possibly
gauginos~\cite{alt}, which modify the radiative corrections and allow
an acceptable \mh.

\item The NuTeV collaboration at Fermilab~\cite{Nutev} have
reported the results of their deep inelastic measurements of
$\frac{\snu_\mu N \RA \snu_\mu X}{\snu_\mu N \RA \mu^{\mp}X}$
using their sign-selected beam. They are able to greatly reduce the
uncertainty in the charm quark threshold in the charged current denominator
by taking appropriate combinations of $\nu_\mu$ and $\bar{\nu}_\mu$.
They find a value for the on-shell weak angle $s_W^2$ of
0.2277(16), which is 3.0$\sigma$ above the global fit value of 0.2228(4).
The corresponding values for the left and right handed neutral current
couplings~\cite{pdg} are $g_L^2 = 0.3001(14)$ and
$g_R^2 = 0.0308(11)$, which are respectively 2.9$\sigma$ below and
0.7$\sigma$ above the expected 0.3040(2) and 0.0300(0).
Possible standard model explanations include an unexpectedly large
violation of isospin in the quark sea~\cite{Nutev}; an
asymmetric strange sea~\cite{davidson}, though NuTeV's data seems to favor the wrong 
sign for this effect;  nuclear shadowing effects~\cite{shadow}; or next to leading
order QCD effects~\cite{davidson}. 

More exotic interpretations could
include a heavy $Z'$ boson~\cite{je,davidson}, although
the standard GUT-type $Z's$ do not significantly improve the fits,
suggesting the need for a $Z'$ with ``designer'' couplings. Mixing
of the $\nu_\mu$ with a heavy neutrino could  account for
the effect~\cite{pati,takeuchi}, and also  for the slightly
low value for the number of light neutrinos $N_\nu = 2.986(7)$
from the $Z$ line shape when $N_\nu$ is allowed to deviate from 3 (this
shows up as a slightly high hadronic peak cross section in the standard
model fit with $N_\nu=3$)~\cite{pdg,je}. This mixing would also
affect muon decay, leading to an apparent Fermi constant smaller than the
true value. This would be problematic for the other $Z$-pole 
observables, but could be compensated by a large negative $T$
parameter~\cite{takeuchi}. However, such mixings would also lead
to a lower value for $|V_{ud}|$, significantly aggravating the
universality problem discussed below.

\item The Brookhaven $g-2$ experiment has reported a precise
new value~\cite{gmin2} using positive muons, leading to a new world average
$a_\mu = 11659203(8) \x 10^{-10}$. Improvements in the statistical
error from negative muon runs are anticipated.
Using
the theoretical value quoted by the experimenters for the
hadronic vacuum polarization contribution,
there is now a small
discrepancy, with the experimental $a_\mu$ larger than the standard model
expectation by $(26 \pm 11) \x 10^{-10}$, a 2.6$\sigma$ effect.
The value and uncertainty in this vacuum polarization are still
controversial\footnote{There are
also uncertainties in the smaller hadronic light by light
diagram. An unfortunate sign error increased the apparent  discrepancy with
the experimental value at an earlier stage, but this has now been corrected.},
so it is hard to know how seriously to take the discrepancy.
One obvious candidate for a new physics explanation would be supersymmetry~\cite{wjm}
with relatively low masses for the relevant sparticles and high $\tan \beta$
(roughly, one requires an effective mass scale of $\tilde{m} \sim 55 \ {\rm GeV} \
\sqrt{\tan \beta}$). There is a correlation between
the theoretical uncertainty in the vacuum polarization and in the
hadronic contribution to the running of $\alpha$ to the $Z$ pole~\cite{corr},
leading to a slight reduction in the predicted Higgs mass when $a_\mu$
is included in the global fit assuming the standard model.

\item 
A few years ago there was an apparent  2.3$\sigma$ discrepancy
between the measured value of the effective  (parity-violating) weak
charge $Q_W(Cs)$ measured in cesium~\cite{apv}, and the expected value.
In particular, cesium has a single electron outside a tightly bound core, 
so the atomic matrix elements could be reliably calculated, leading
(it was thought)
to a combined theoretical and experimental uncertainly of around 0.6\%.
It was subsequently pointed out, however, that there was a significant
contribution from the Breit (magnetic) interaction between two
electrons~\cite{apv0}, reducing the discrepancy. Further calculations
of the $O(Z \alpha^2)$ radiative corrections associated with the
vacuum polarization in the nuclear Coulomb field  reinstated
 the discrepancy, yielding 
$Q_W = -72.12(28)(34)$~\cite{apv1} and $-72.18(29)(36)$\cite{apv2}, which
are respectively 2.2 and 2.0$\sigma$ above the expected $-73.09(3)$.
An even more recent calculation of the electron line corrections~\cite{apv2}
(including some estimated higher order effects) was surprising large,
yielding $-72.83(29)(39)$, in agreement with the standard model.
The situation is confusing, and it is not certain whether the last
word has been written, but clearly at this point there is no
evidence for new physics.

\item 
The unitarity of the CKM matrix can be partially tested
by the universality prediction that 
$\Delta \equiv 1 - |V_{ud}|^2 - |V_{us}|^2 - |V_{ub}|^2 $
should vanish. In particular $|V_{ud}|$ can be determined
by the ratio of $G^V_\beta/G_\mu$, where $G^V_\beta$ and
$G_\mu$ are respectively the vector coupling in $\beta$ decay
and the $\mu$ decay constant. The most precise determination
of $|V_{ud}|$ is from superallowed $0^+ \RA 0^+$ transitions,
currently yielding $|V_{ud}| = 0.9740(5)$~\cite{superallowed}.
Combining with $|V_{us}|$ from kaon and hyperon decays and
$|V_{ub}|$ from $b$ decays, this yields 
a 2.3$\sigma$ discrepancy $\Delta = 0.0032(14)$. It is unlikely
that the uncertainties in $|V_{us}|$ or $|V_{ub}|$ could be
responsible, suggesting either the presence of unaccounted-for new
physics, or, possibly, effects from  higher order isospin violation such
as nuclear overlap corrections.
However, the latter have been
carefully studied, so the effect may be real.
This problem has been around
for some time, but until recently less precise determinations from
neutron decay were consistent with universality.
Recently, a more precise measurement of the neutron decay
asymmetry has been made by the PERKEO-II group at ILL~\cite{neutron}.
When combined with the accurately known neutron lifetime, this allowed
the new determination $|V_{ud}|= 0.9713(13)$, implying
$\Delta = 0.0083(28)$, i.e., a 3$\sigma$ violation of unitarity. Note,
however, that this value is only marginally consistent with the value
obtained from superallowed transitions.

Mixing of the $\nu_\mu$ with a heavy neutrino, suggested as a solution
of the NuTeV anomaly, would mean that $G_\mu$ is larger than the apparent value
and would aggravate this discrepancy. ($\nu_e$ mixing would affect
$G^V_\beta$ and $G_\mu$ in the same way and have no effect.)
On the
other hand, a very small mixing of the $W$ boson with a heavy $W'$ coupling
to right handed currents, as in left-right symmetric models,
could easily account for the discrepancy for the appropriate sign for the
mixing~\cite{lr}, especially if the right-handed neutrinos are Majorana
and too heavy to be produced in the decays.

\item 
The LEP and SLC $Z$-pole experiments are the most precise tests of the
standard electroweak theory, but they are blind to any new
physics that doesn't affect the $Z$ or its couplings. Non-$Z$-pole
experiments are therefore extremely important, especially given
the possible NuTeV anomaly. In the near future we can expect
new results in polarized M\o ller scattering from SLAC~\cite{moller}, and
in the QWEAK polarized electron experiment at Jefferson Lab~\cite{qweak}.
\end{itemize}

\section{Fit Results}

As of June, 2002, the result of the global fit is~\cite{je}

\bqa
           M_H &=& 86^{+49}_{-32} \mbox{ GeV}, \nonumber \\
           m_t &=& 174.2  \pm 4.4  \mbox{ GeV}, \nonumber  \\
      \alpha_s &=& 0.1210 \pm 0.0018, \nonumber \\
  \hat{\alpha}(M_Z)^{-1} & = & 127.922 \pm 0.020 \nonumber \\
   \hat{s}^2_Z &=& 0.23110 \pm 0.00015, \nonumber \\
  \bar{s}^2_\ell &=& 0.23139 \pm 0.00015, \nonumber \\
          s^2_W &=& 0.22277 \pm 0.00035 \nonumber \\
          s^2_{M_Z} &=& 0.23105 \pm 0.00008 \nonumber \\
\chi^2/{\rm d.o.f.} & = & 49.0/40 (15\%) \label{results}
\eqa
 This is in generally good agreement with the fit of the LEP Electroweak
Working Group of May, 2002~\cite{lepewwg},
\bqa
 M_H &=& 85^{+54}_{-34}\mbox{ GeV}, \nonumber \\
\alpha_s &= &0.1183 \pm 0.0027, \nonumber \\
m_t &=&174.7^{+4.5}_{-4.3} \mbox{ GeV} \nonumber \\
\bar{s}^2_\ell &=& 0.23137 \pm 0.00015, \nonumber \\
s_W^2 &=& 0.22272 \pm 0.00036, \nonumber \\
\eqa
 up  to understood effects associated
with the input data set, higher order terms, and the value of \delhad.
The agreement is especially impressive since the LEPEWWG 
analysis utilized expressions computed in the
on-shell scheme, while the analysis in (\ref{results}) used \msb.
The slightly higher value  \als $=0.1210 \pm 0.0018$ in (\ref{results}) 
is due in part to the inclusion of the new $\tau$ lifetime result~\cite{tauwidth,je}.
(Without it, one would obtain \als=$0.1200 \pm 0.0028$.) These
values for \als \ are in reasonable agreement with the previous world average
$\als = 
0.1172(20)$, which includes other determinations, most of which
are dominated by theoretical uncertainties~\cite{hinchliffe}. 
The $Z$-pole value is insensitive to oblique new physics, but is
very sensitive to non-universal new physics, such as those which affect
the $Z b \bar{b}$  vertex.

\begin{figure}[h]
\centering
\includegraphics*[scale=0.6]{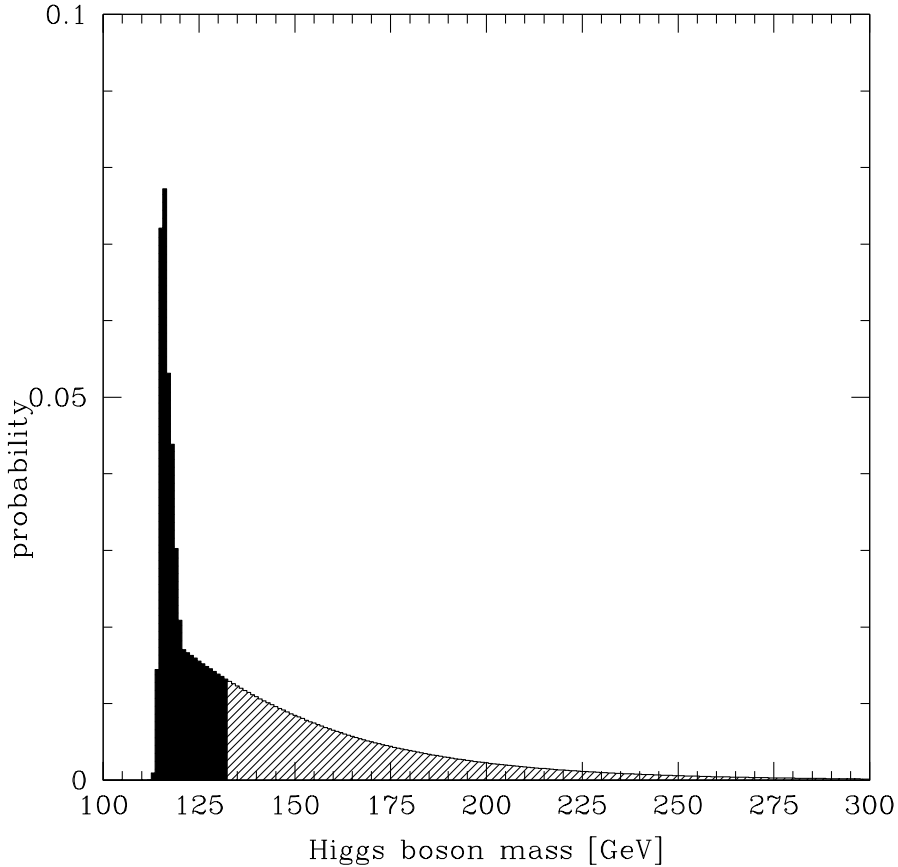}
\caption{Allowed regions in \mh \  vs \mt \ from precision data,
compared with the direct exclusion limits from LEP2. Courtesy of Jens Erler.}
\label{higgspdf}
\end{figure}

The prediction for the Higgs mass from indirect data, 
\mh $= 86^{+49}_{-32}$ GeV, should be compared with the 
direct LEP2 limit~\cite{lephiggs} $\mh \simgr 114.4 (95\%)$ GeV.
The theoretical range in the standard model is
 115 GeV $\simle \mh \simle$ 750 GeV,
where the lower (upper) bound is from vacuum stability (triviality).
In the MSSM, one has
 $\mh \simle 130$ GeV, while \mh \ can be as high as 150 GeV in generalizations.
 Including the direct LEP2 exclusion results, one finds
$\mh < 215$ GeV at 95\%. The probability distribution for \mh,
including both direct and indirect constraints and
updated from the analysis in~\cite{erler}\footnote{See also
the study in~\cite{degrassi}},
 is shown in Figure \ref{higgspdf}.

\mh \ enters the expressions for the radiative corrections logarithmally. It is fairly 
robust to many types of new physics, with some exceptions. In particular,
a much larger \mh \
 would be allowed for
 negative values for the $S$ parameter or positive values for $T$.
The predicted value would decrease if new physics accounted for
the value of $A_{FB}(b)$~\cite{chanowitz}.

\section{Beyond the Standard Model}
The 
$\rho_0$ or $S$, $T,$ and $ U$ parameters describe the tree level effects of
Higgs triplets, or the loop effects on the $W$ and $Z$ propagators due
to such new physics as
nondegenerate fermions or scalars, or chiral families (expected,
for example,  in extended technicolor). The current values are:
 \bqa  S &=& -0.14 \pm 0.10 (-0.08)  \nonumber  \\
T &=& -0.15 \pm 0.12 (+0.09)  \nonumber \\
U &=& 0.32 \pm 0.12 (+0.01)  \ \ \  (2.6\sigma) \nonumber
\eqa
for $M_H = 115.6 \ (300)$ GeV,
where these represent the effects of new physics only (the \mt \ and \mh \ effects
are treated separately).
Similarly,
 $\rho_0  \sim 1 + \alpha T = 0.9997^{+0.0011}_{-0.0008}$
for $M_H = 73^{+106}_{-34}$
GeV and $S=U=0$. If one constrains
$T = U =0$, then $S=0.10^{+0.12}_{-0.30}$. There is a strong negative $S-\mh$
correlation, so that the Higgs mass constraint is relaxed to $M_H < 570$
GeV at 95\%. 
For \mh \ fixed at 115.6 GeV, one finds
$S = -0.040(62)$, which implies that the number of ordinary plus
degenerate heavy families
is constrained to be
$N_{\rm fam} = 2.81 \pm 0.29$.  This
is
complementary to the lineshape constraint,  
$N_\nu = 2.986 \pm 0.007 $, which  only applies  to neutrinos less massive
than $M_Z/2$. One can also restrict   additional nondegenerate families by allowing
both $S$ and $T$ to be nonzero, yielding
$N_{\rm fam} = 2.79 \pm 0.43$  for $T = -0.01 \pm 0.11$.


\begin{thebibliography}{99}
\bibitem{pdg} See the review
{\it Electroweak model and constraints on new physics},
J. Erler and P. Langacker, 
in the 2002 Review of Particle Physics,
K. Hagiwara et al., Phys. Rev. D66, 010001 (2002).
Older references and reviews are cited theirin.

\bibitem{lepewwg}
Preliminary results of the LEP Electroweak Working Group
 as of May 2002 may be found in
The LEP Collaborations, LEPEWWG/2002-01, available at
http://lepewwg.web.cern.ch/LEPEWWG/.

\bibitem{snowmass}
P.~Langacker,
in {\it Proc. of the APS/DPF/DPB Summer Study on the Future
of Particle Physics (Snowmass 2001)}, ed. N.~Graf,
http://www.slac.stanford.edu/econf/C010630/,
hep-ph/0110129.

 
\bibitem{je} Jens Erler, private communication.



 \bibitem{lephiggs}
Preliminary results, as of July 2002, may be found in
the LEP Higgs Working Group report  LHWG Note/2002-01,
available at http://lephiggs.web.cern.ch/LEPHIGGS.

\bibitem{ALEPH}
A.~Heister {\it et al.}  [ALEPH Collaboration],
Phys.\ Lett.\ B {\bf 526}, 191 (2002).

\bibitem{tauwidth}
J.~Erler and M.~x.~Luo,
hep-ph/0207114.

\bibitem{pl}
P. Langacker, these proceedings.

\bibitem{zpr}
J.~Erler and P.~Langacker,
Phys.\ Rev.\ Lett.\  {\bf 84}, 212 (2000).

\bibitem{fcnc}
P.~Langacker and M.~Pl\"umacher,
Phys.\ Rev.\ D {\bf 62}, 013006 (2000).

\bibitem{wagner}
D.~Choudhury, T.~M.~Tait and C.~E.~Wagner,
Phys.\ Rev.\ D {\bf 65}, 053002 (2002).

\bibitem{chanowitz} 
M.~S.~Chanowitz,
Phys.\ Rev.\ Lett.\  {\bf 87}, 231802 (2001);
Phys.\ Rev.\ D {\bf 66}, 073002 (2002).

\bibitem{alt}
G.~Altarelli, F.~Caravaglios, G.~F.~Giudice, P.~Gambino and G.~Ridolfi,
JHEP {\bf 0106}, 018 (2001).

\bibitem{Nutev}
G.~P.~Zeller {\it et al.}  [NuTeV Collaboration],
Phys.\ Rev.\ Lett.\  {\bf 88}, 091802 (2002);
Phys.\ Rev.\ D {\bf 65}, 111103 (2002);
hep-ex/0207037, 0207052, 0210010;
R.~H.~Bernstein,
hep-ex/0210061.

\bibitem{davidson}
S.~Davidson, S.~Forte, P.~Gambino, N.~Rius and A.~Strumia,
JHEP {\bf 0202}, 037 (2002);
S.~Davidson,
arXiv:hep-ph/0209316.

\bibitem{shadow}
G.~A.~Miller and A.~W.~Thomas,
hep-ex/0204007;
W.~Melnitchouk and A.~W.~Thomas,
hep-ex/0208016;
S.~Kovalenko, I.~Schmidt and J.~J.~Yang,
Phys.\ Lett.\ B {\bf 546}, 68 (2002);
S.~Kumano,
hep-ph/0209200. Some of these papers are commented on in~\cite{Nutev}.


\bibitem{pati}
K.~S.~Babu and J.~C.~Pati,
hep-ph/0203029.

\bibitem{takeuchi}
W.~Loinaz, N.~Okamura, T.~Takeuchi and L.~C.~Wijewardhana,
hep-ph/0210193.



\bibitem{gmin2}
G.~W.~Bennett {\it et al.}  [Muon g-2 Collaboration],
Phys.\ Rev.\ Lett.\  {\bf 89}, 101804 (2002)
[Erratum-ibid.\  {\bf 89}, 129903 (2002)].




\bibitem{wjm} See, for example,
A.~Czarnecki and W.~J.~Marciano,
Phys.\ Rev.\ D {\bf 64}, 013014 (2001).


\bibitem{corr}
J.~Erler and M.~x.~Luo,
Phys.\ Rev.\ Lett.\  {\bf 87}, 071804 (2001).


%

\bibitem{apv}
S.C. Bennett and C.E. Wieman, { Phys. Rev. Lett.} {\bf 82}, 2484
(1999).

\bibitem{apv0}
A.~Derevianko,
Phys.\ Rev.\ Lett.\  {\bf 85}, 1618 (2000).

\bibitem{apv1}
W.~R.~Johnson, I.~Bednyakov and G.~Soff,
Phys.\ Rev.\ Lett.\  {\bf 87}, 233001 (2001)
[Erratum-ibid.\  {\bf 88}, 079903 (2002)].

\bibitem{apv2}
V.~A.~Dzuba, V.~V.~Flambaum and J.~S.~Ginges,
hep-ph/0204134.

\bibitem{apv3}
M.~Y.~Kuchiev,
hep-ph/0208196.


\bibitem{superallowed}
I.~S.~Towner and J.~C.~Hardy,
Phys.\ Rev.\ C {\bf 66}, 035501 (2002).

\bibitem{neutron}
H.~Abele {\it et al.},
Phys.\ Rev.\ Lett.\  {\bf 88}, 211801 (2002).

\bibitem{lr} See the articles by 
J. Deutsch and P. Quin, p. 706, and by A. Sirlin, p. 766, in
{\it Precision Tests of the Standard Electroweak Model}, ed.
P. Langacker (World, Singapore, 1995).


\bibitem{moller}
SLAC E158: http://www.slac.stanford.edu/exp/e158/.

\bibitem{qweak} The QWEAK experiment,
http://www.jlab.org/qweak/.

\bibitem{hinchliffe}
See the review
{\it Quantum chromodynamics},
I. Hinchliffe,
in the 2002 Review of Particle Physics.

\bibitem{erler}
J.~Erler,
Phys.\ Rev.\ D {\bf 63}, 071301 (2001) and these proceedings.

\bibitem{degrassi}
G. Degrassi, these proceedings.




\bibitem{ack}
It is a pleasure to thank Jens Erler for many of the results
in this update.
This work was supported by the U.S. Department
of Energy grant DOE-EY-76-02-3071.
\end{thebibliography}
\end{document}